\documentclass[12pt,preprint]{aastex}

\usepackage{emulateapj5,apjfonts}

\slugcomment{Submitted to ApJ Letters, \today}
\shorttitle{THERMALLY DRIVEN ISM TURBULENCE}
\shortauthors{KRITSUK \& NORMAN}

\begin{document}
\def\lsim{~\raise0.3ex\hbox{$<$}\kern-0.75em{\lower0.65ex\hbox{$\sim$}}~}
\def\gsim{~\raise0.3ex\hbox{$>$}\kern-0.75em{\lower0.65ex\hbox{$\sim$}}~}

\journalinfo{The Astrophysical Journal, 580:L000-L000, November 20, 2002}

\submitted{Received 2002 June 14; accepted 2002 October 3}
\title{Interstellar Phase Transitions Stimulated by Time-dependent Heating}

\author{Alexei G. Kritsuk and Michael L. Norman}
\affil{Department of Physics and Center for Astrophysics and Space Sciences, 
University of California at San Diego,\\
9500 Gilman Drive, La Jolla, CA 92093-0424; 
akritsuk@ucsd.edu, mnorman@cosmos.ucsd.edu}

\begin{abstract}
We use three-dimensional hydrodynamic numerical simulations to study phase 
transformations occurring in a clumpy interstellar gas exposed to 
time-dependent volumetric heating.
To mimic conditions in the Galactic interstellar medium, 
we take a numerical model of a turbulent multiphase medium from Kritsuk
\& Norman (2002) computed in a periodic box with mean density 
$n_0=0.25$~cm$^{-3}$ and mean pressure $P_0/k\sim10^{3.4}$~K~cm$^{-3}$. 
A second model with $n_0=1$~cm$^{-3}$ is also considered.
Variations of the heating rate on a timescale of 1--10~Myr applied 
thereafter cause pressure variations in the gas and shifting of the thermal 
equilibrium curve in the phase plane.
This stimulates mass transfer between the gas phases via thermal 
instability, converting 5--10~\% of the thermal energy into kinetic 
energy of gas motions.
The experiments demonstrate that recurrent substantial heating episodes 
can maintain turbulence at this level.
Possible applications to the interstellar gas heated by
variable far-ultraviolet background radiation produced by short-living 
massive stars are discussed.
\end{abstract}


\section{Introduction}
Thermal instability (TI), controlled by the interplay of volumetric
heating and radiative cooling of optically thin gas, is one of the key 
physical processes operating in the interstellar medium (ISM) 
\citep{field65,meerson96}.
For a long time it was believed that an important manifestation of TI is 
the splitting of the ISM into multiple thermal phases that coexist in pressure
equilibrium \citep{pikel'ner68,field..69,mckee.77,mckee90,heiles01a}.
Recent numerical studies extended this classical picture, showing that 
interstellar turbulence quite naturally appears as a by-product of the same 
phase transformations that provide the irregular, clumpy and filamentary 
density substructure in the ISM \citep[hereafter KN02]{koyama.02,kritsuk.02}.
In the absence of continuous driving, however, this turbulence decays
as $t^{-\eta}$, with $1\lsim\eta\lsim2$ 
[KN02; cf. \citet{maclow...98,stone..98}], 
leaving behind a ``fossil'' isobaric density distribution with no 
substantial velocity structure remaining at the scales of the 
density inhomogeneities [cf. \citet{mccray..72}].

Since observations indicate the persistence of turbulence and substructure
in the ISM on a variety of timescales and length scales, 
there must be some source of energy for turbulence support.
A number of hydrodynamic and MHD mechanisms have been 
proposed to sustain turbulent motions within molecular clouds and
other thermal phases of the ISM.
These include winds from young stars \citep{norman.80,franco.84}, 
photoionization-regulated star formation \citep{mckee89}, 
supernova explosions \citep{mckee.77,korpi...99,kim..01}, largescale
external shocks \citep{kornreich.00}, and 
differential rotation \citep{richard.99,sellwood.99,wada.99}.
They rely on a combination of energy deposition from star-forming 
activity, self-gravity, magnetic effects, or galactic rotation to feed
the turbulence.
These options do not exhaust all possible driving mechanisms,
cf. \citet{norman.96}.
Motivated by the results of KN02 on TI-induced ISM turbulence, we study here 
the possibility of purely ``thermal'' support for interstellar turbulence.
This mechanism, which is physically distinct from the local mechanisms listed
above, can operate {\em in addition} to them.

Most previous works focused specifically on TI implicitly assume the 
existence of stationary thermodynamic equilibrium.
The ISM, however, is highly time-dependent \citep{gerola..74,bania.80}.
In particular, time variations can be caused by changes in the level of 
ISM heating.
The main source of energy input for the neutral gas is background radiation 
in the far-ultraviolet (FUV) part of the spectrum \citep{wolfire....95}.
The FUV field is generally due to OB associations of quickly evolving
massive stars that form in giant molecular clouds.
For example, in the local ISM the expected FUV energy density undergoes 
substantial fluctuations on a wide range of timescales from $\lsim10$ to 
$\sim100$~Myr \citep{parravano..02}.
The focus of the numerical simulations presented in this Letter is on 
the hydrodynamic effects of the time-dependent heating of the ISM.

\begin{figure*}
\epsscale{2.0}
\plotone{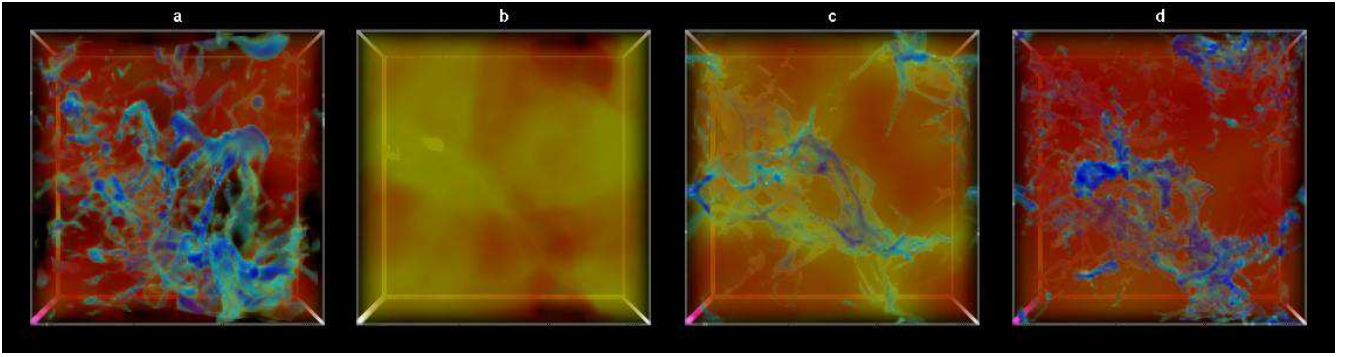}
\vspace{0.3cm}
\caption{Snapshots of the gas density field (perspective volume rendering):
({\em a}) Turbulent multiphase gas at $t=2$~Myr;
({\em b}) relaxed state at a high heating rate with no cold phase present, 
$t=2.8$~Myr;
({\em c}) violent relaxation to an equilibrium with reduced heating, 
(the seeds for the new population of cold clouds are forming along caustics in
the stable warm phase, $t=3.24$~Myr);
({\em d}) partially relaxed state at a low heating rate, $t=4$~Myr
(20~pc box, $128^3$ grid points).
The log density color coding is as follows: 
The most dense blobs, $n > 8$~cm$^{-3}$, are blue;
the less dense gas, $n\in[3, 8]$~cm$^{-3}$, is light blue to green;
the warm gas at $n\in[0.5, 3]$~cm$^{-3}$ is yellow to red, 
and the low-density gas ($n < 0.5$~cm$^{-3}$) is transparent.
The figure is also available as an mpeg animation in the electronic 
edition of the {\em Astrophysical Journal}.
\label{fig1}} 
\end{figure*}

\section{Thermal forcing for interstellar turbulence}

\subsection{Numerical Experiments}

We solve the equations of ideal gasdynamics (eqs. [6]-[9] in Field 1965) in
a cubic domain with periodic boundary conditions, assuming zero conductivity
and no gravity.
More detailed information regarding the simulations' setup and numerical 
techniques used can be found in KN02.
Each of our numerical solutions represents a one-parameter family of models
in physical space with a box size $\xi L$, a mean gas density in
the box $\xi^{-1}\rho_0$, and a heating rate $\xi^{-1}\Gamma$.
Other variables should be scaled accordingly, e.g., time scales as $\xi t$.
The dimensionless factor $\xi$ resembles scaling properties of the governing 
equations.

\subsection{Thermal Phase Fractions Versus Mean Gas Density}

In order to study the sensitivity of emerging thermal phases to our model 
parameters, we performed a series of low-resolution ($128^3$) simulations 
with a fixed value of the heating rate $\Gamma$, varying the initial gas 
temperature, mean gas density $\rho_0$, and box size $L$.
As in KN02, we conventionally define the thermal phases in terms of 
the equilibrium gas temperature based on the \citet{field65} TI criterion 
applied to ${\mathcal L}(\rho,T)=\rho\Lambda(T) - \Gamma$.
The cold stable phase has $T<600$~K (phase H), 
the unstable regime falls into the range $T\in[600, 8000]$~K (phase G), 
and the warm stable phase has $T\in[8000, 19,000]$~K (phase F).
We refer to the transient regime with $T>19,000$~K (dependent on initial
conditions) as the ``hot ionized medium'' (hereafter HIM).

The mass fractions of the thermal phases ($f_F$, $f_G$, and $f_H$) in a relaxed
state mostly depend on the mean gas density in the box $\rho_0$.
If $\rho_0$ is low enough, both the unstable and cold phases are transient,
and evolution ends up with a single warm gas phase once the
turbulence generated during the violent relaxation to thermal equilibrium
decays.
If $\rho_0$ is higher than $\rho_{min}$, corresponding to the local pressure
minimum on the bistable thermal equilibrium curve, the evolution ends 
up with a single cold phase.
For intermediate mean gas densities, the system evolves to a multiphase state
with an asymptotic $f_H/f_F$ ratio dependent solely on the value of 
$\rho_0$.\footnote{Neither the initial temperature nor the box size will 
modify 
$f_H/f_F$, provided that isobaric modes of TI are resolved in the unstable
density regime in the vicinity of thermal equilibrium.
Unresolved TI results tend to underestimate the cold gas fraction, 
overestimate the unstable gas fraction, and underestimate the turbulence
decay rate [returning $\eta\approx1$ rather than $\eta\sim 2$, see
A.~G.~Kritsuk \& M.~L.~Norman (2002, in preparation)].}
The mass fraction of thermally unstable gas can be $\gsim50$\%
during the violent relaxation stage (including the cases with transient cold 
phase), but decreases to $\lsim$5\% later, when it also strongly correlates 
with the rms Mach number ($f_G\propto{\mathcal M}^{\beta}$, 
where $\beta\sim{2\over3}$).

\begin{figure*}
\epsscale{1.0}
\plotone{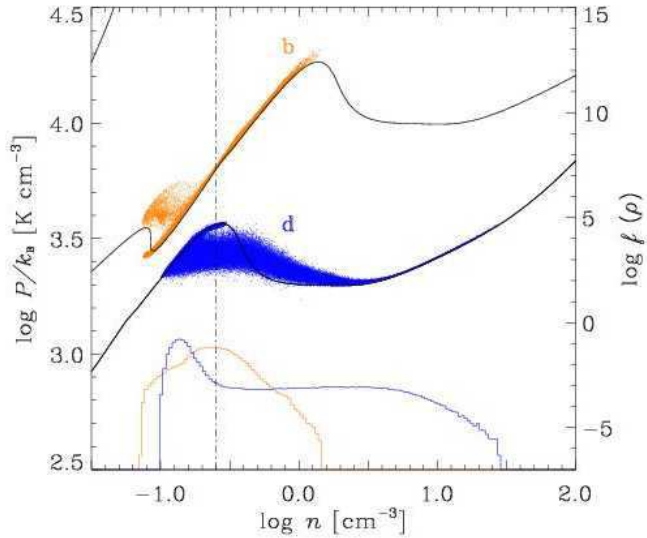}
\caption{Scatter plots of gas pressure vs. gas density at 
$t=2.88$~Myr and $4.0$~Myr (snapshots {\em b} and {\em d} in Fig.~\ref{fig1}).
The solid lines show thermal equilibria at high and low heating states.
The dash-dotted line shows the mean gas density.
Corresponding density PDFs are plotted at the bottom (see scale to the right).
\label{fig2}} 
\end{figure*}

\subsection{Variable Heating Rate and Thermal Forcing}

Time variations in the heating rate $\Gamma$ translate the 
location of the thermal equilibrium curve ${\mathcal L}(\rho,T)=0$ in the 
phase plane $(\rho, P)$ along an isotherm 
(leaving the shape of the curve unchanged if $\Gamma=const$).
For a given radiative cooling efficiency, the higher the heating rate, the 
higher the equilibrium pressure and density, 
in particular, $\rho_{min}\propto\Gamma$.
Thus, global $\Gamma$ variations in a medium with constant mean 
density on a sufficiently short timescale will necessarily entail 
some phase adjustment.
Oscillations of the gas pressure and redistribution of mass between the 
thermal phases generate gas flows that gain their kinetic energy 
at the expense of the thermal energy supplied by the heating source.
Phase restructuring in response to variations of $\Gamma$ involves
a broad range of length scales and produces highly irregular flows since the
density distribution in a multiphase medium itself is irregular.
Therefore, to a certain extent, turbulence can be sustained by time-dependent 
heating if substantial variations occur on a timescale that is not too long 
compared with the turbulence decay timescale.
Otherwise, during long periods of quiescence, the turbulence would get 
fossilized.

In the following, we present results for two models: 
(1) a low-$\rho_0$ case in which the cold phase gets fully dissolved into the
warm phase during the high heating state and in which the medium hovers 
between single- and two-phase states as $\Gamma$ varies, and 
(2) an intermediate-$\rho_0$ case with varying phase content in a two-phase
medium.

\subsection{Low-density Medium}

For our low-density run, we take initial conditions from the fiducial model of 
KN02,\footnote{We rescale physical units of the KN02 low-resolution 
($128^3$ grid zones) fiducial model to better fit local interstellar 
conditions ($\xi=4$, time: $t = 4t_{KN}$, box size: $L=4L_{KN}=20$~pc,
mean gas density: $\rho_0=\rho_{KN}/4=0.25 m_H$~g~cm$^{-3}$, 
pressure: $p=p_{KN}/4$, and heating rate: 
$\Gamma_{low}=\Gamma_{KN}/4$).} at $t=2$~Myr (Fig. \ref{fig1}{\em a}).
By this time, the gas has partially relaxed to a bistable thermal
equilibrium after violent radiative cooling from a temperature of 
$2\times10^6$~K.
However, the gas pressure variations still remain substantial
($\Delta P/P\sim3$, see Fig. 2{\em e} in KN02 for a phase diagram) and 
$\mathcal{M}$ is slightly above one.
For average conditions at $t=2$~Myr in the thermally unstable regime, the 
cooling scale is $\lambda_p\sim4$~pc and the conductive (Field) scale is 
$\lambda_{\kappa}\sim 0.4$~pc (see KN02 for definitions).
Thus, on our low resolution grid, the cooling scale is marginally resolved, 
and the conductive scale falls below the resolution limit.

The higher resolution model ($256^3$) of KN02 at $t=2$~Myr returns a 
slightly higher value for $f_H$ and slightly lower value for $f_G$, while 
$f_F$ is roughly the same as in the low-resolution model
(the relative differences are 0.16, 0.16, and 0.005, respectively).
The relative difference in generated kinetic energy for the high- and 
low-resolution models at $t=2$~Myr is 0.013.
At later times, as relaxation becomes less violent, the deviations get 
smaller.
We consider the above values to be good upper estimates for the
effects of numerical diffusion.
Therefore, the conclusions that we make below regarding the 
phase fractions and 
energy conversion efficiency are robust within the specified limits.

\begin{figure*}
\epsscale{2.0}
\plottwo{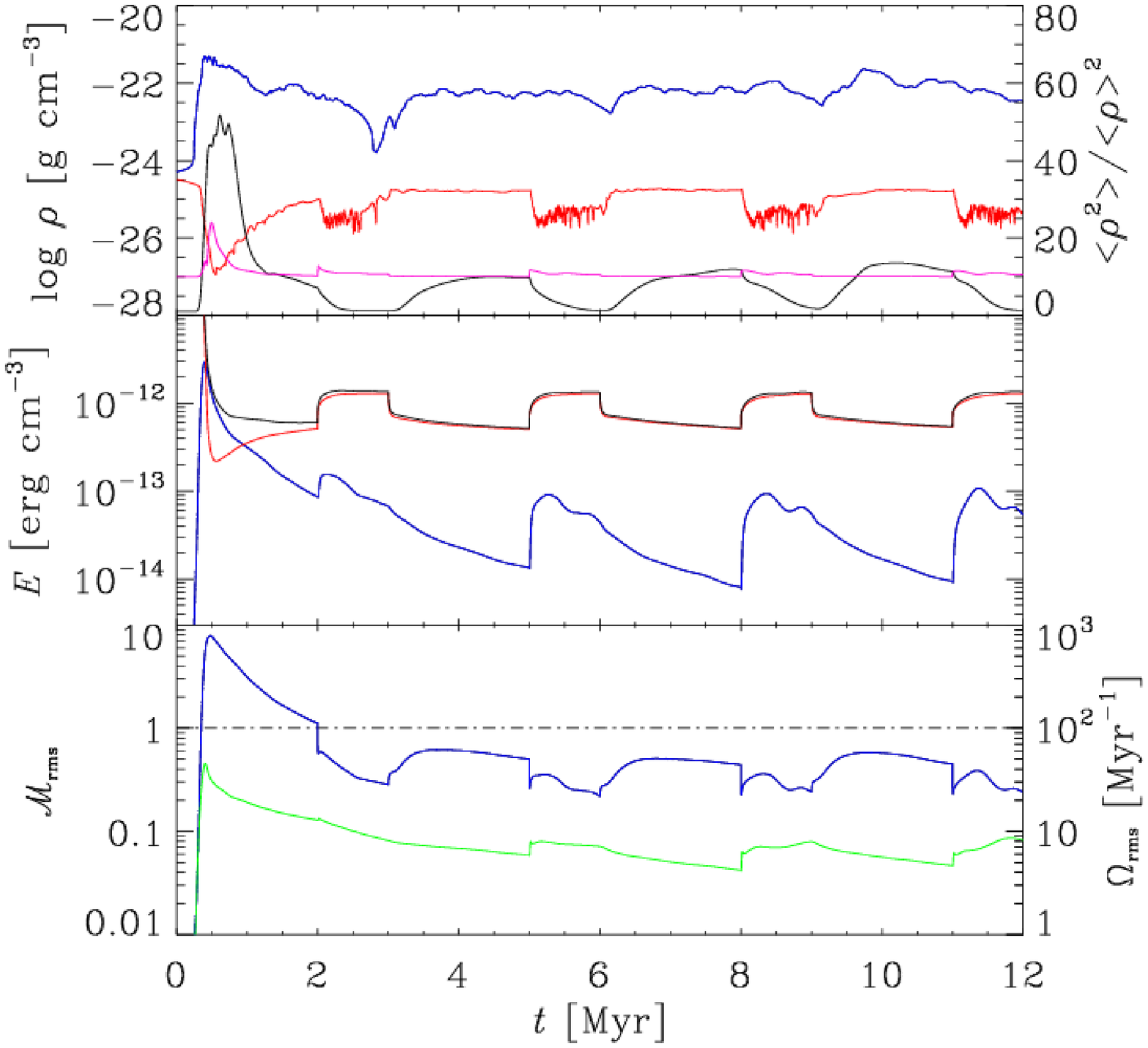}{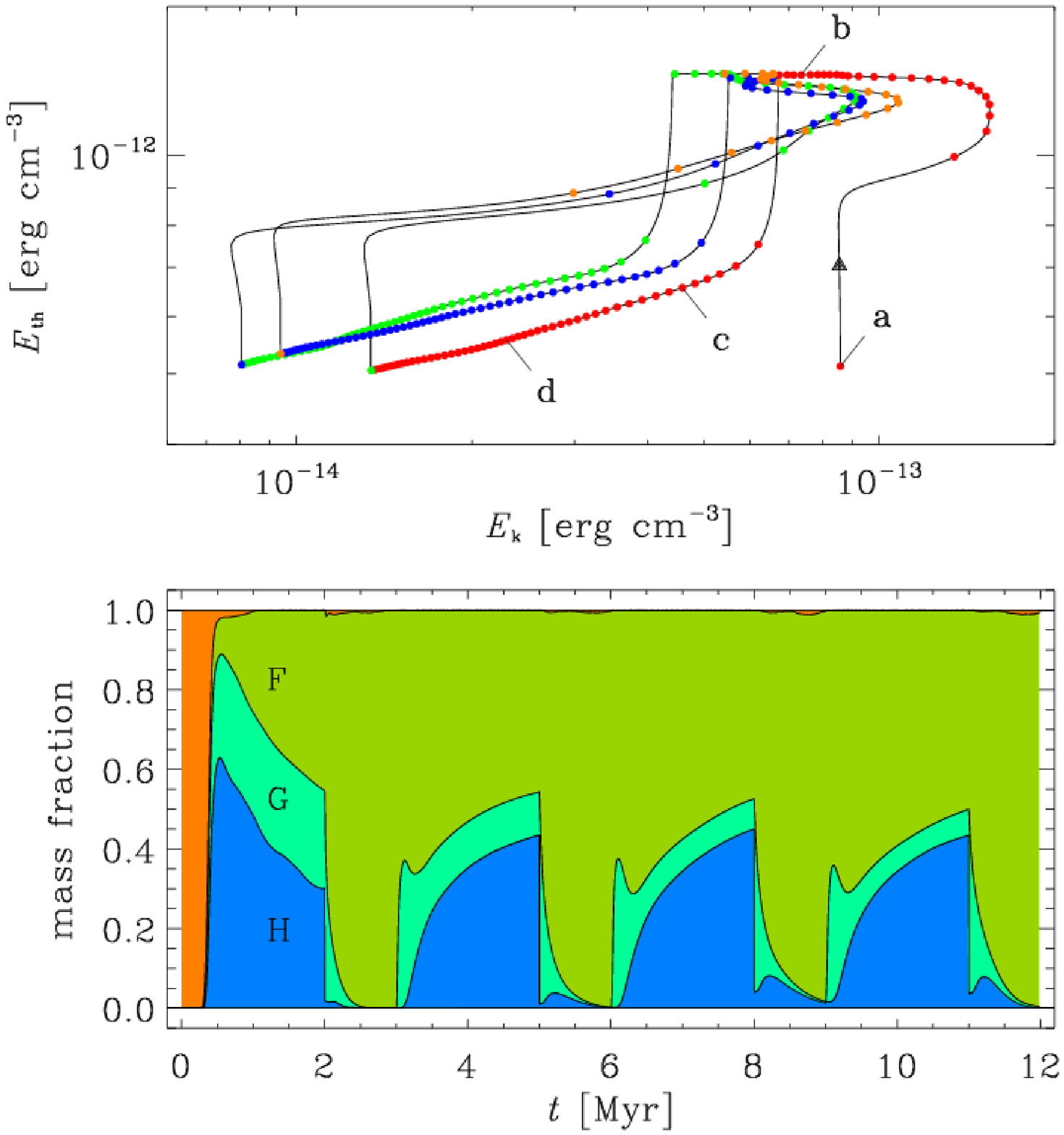}
\caption{Time evolution of global variables. {\em Top left panel}: 
$\rho_{max}$ ({\em blue}),
$\rho_{min}$ ({\em red}), $\langle\rho^2\rangle/\rho_0^2$ ({\em black}), 
$10\langle P^2\rangle/\langle P\rangle^2$ ({\em magenta},
scale is at the right).
{\em Middle left panel}: Total energy ({\em black}), 
thermal energy ({\em red}), kinetic
energy ({\em blue}).
{\em Bottom left panel}: Mass-weighted rms Mach number ({\em blue}), 
rms enstrophy ({\em green}).
{\em Top right panel}: Mean thermal energy vs. mean kinetic energy for a 
time interval $t\in[2, 12]$~Myr covering 
3.5 periods (the arrow shows the direction of evolution).
The trajectory is marked up with filled circles separated by 0.04~Myr 
(color changes from one period to another).
Labels {\bf a}, {\bf b}, {\bf c}, and {\bf d} indicate positions corresponding
to the four frames in Fig.~\ref{fig1}.
{\em Bottom right panel}: Mass fractions of thermal phases:
warm stable (F), intermediate unstable (G), and cold stable (H).
\label{fig3}} 
\end{figure*}

At $t=2$~Myr, we apply enhanced heating rate $\Gamma_{high}=5\Gamma_{low}$ 
for a span of 1~Myr and then switch back to $\Gamma_{low}$ and keep it low 
for the next 2~Myr.
We follow the evolution for a few cycles up to $t=12$~Myr using a
piecewise-constant representation for $\Gamma(t)$ with a period of 
3~Myr.
The mean gas density is below the minimum required for coexistence of two 
stable phases in pressure equilibrium during the {\em high state}, 
and it is slightly below the density value that corresponds to the local 
pressure maximum on the thermal equilibrium curve for the {\em low state}.
The results are shown in Figures \ref{fig1}, \ref{fig2}, and \ref{fig3}.

Overall, the evolution consists of a series of phase transitions, which
follow the abrupt changes in the heating rate.
Figure~\ref{fig2} displays the thermal equilibria for $\Gamma_{high}$ 
({\em upper curve}) and $\Gamma_{low}$ ({\em lower curve}) and scatter 
plots of gas pressure versus density for partially relaxed states 
corresponding to the density fields shown in Figure~\ref{fig1}{\em b} 
(high state) and Figure~\ref{fig1}{\em d} (low state).
After the change in heating rate at $t=2$~Myr, gas quickly ($\lsim0.04$~Myr) 
switches to the high-state thermal equilibrium, but it then takes up to 0.7~Myr
to completely dissolve the preexisting cold phase into the warm phase.
A transition to the high state is accompanied by the growth of thermal
and kinetic energy and by the decrease in the rms Mach number and
density variance (Fig. \ref{fig3}, {\em left panels}).
Also, the gas density range shifts to somewhat lower values,
and warm phase gas enters a transonic regime as a result of the large 
pressure variations ($\Delta P/P\sim30$) accompanying the relaxation process. 
The enhanced heating episodes are also efficient in generating vorticity, 
as can be seen from the enstrophy plot in Figure~\ref{fig3}.
The transition back to the low state at $t=3$~Myr is also abrupt.
It entails the formation of a new population of cold ``clouds'' along caustics 
in the warm-phase velocity field.
This mechanism naturally produces dense structures that morphologically
can be described as ``blobby sheets,'' introduced by \citet{heiles.02}.
The seeds of newly forming clouds can be seen in Figure~\ref{fig1}{\em c} and
the clouds' morphology can be traced in the accompanying mpeg animation.
The density probability distribution function (PDF)
undergoes substantial changes during each period of evolution, loosing and
regenerating its high-density shoulder in the high and low heating states, 
respectively (see Fig.~\ref{fig2}).
If a longer relaxation period were allowed in between the sequential high 
states, a bimodal density PDF would have developed.

We ran the simulation for more than three periods to see how the
medium adjusts to periodic variations of the heating rate.
We found that kinetic energy peaks at  $\sim7$~\% of the maximum thermal 
energy in about 0.4~Myr after the enhanced heating is switched on 
(Fig. \ref{fig3}).
This result is insensitive to the initial conditions and to the
particular choice of the low state duration.
The system settles onto an ``attractor'' in the $(E_k, E_{th})$--plane.
The maximum Mach number in the box oscillates in the range between 1.2 and
5 and roughly follows the evolutionary pattern of ${\mathcal M}_{rms}$.
Thus, thermally driven turbulence in a low-density medium is transonic.

There is a tendency for the unstable gas mass fraction, accompanying the
production of the cold phase in the low state, to decrease slowly from cycle 
to cycle.
The average value of $f_G$ is 9.5\% or 11\% for averaging over time intervals 
$t\in[2,12]$~Myr or $t\in[0,12]$~Myr, respectively.
The mean mass fraction of the cold phase $f_H$ in the high states displays 
some secular oscillations, so that the cold phase does not always get 
completely dissolved during the high state.
This means that the value of $\rho_0$, chosen for this model, is close to a
critical value that separates single- and two-phase media heated 
at~$\Gamma_{high}$.
The $f_H$-values grow up to $\sim45$\% while the system remains in the low 
state.
Were the low state to last longer, $f_H$ would approach the asymptotic value 
of $\sim55$\%, recorded for the low-resolution fiducial run of KN02.
The unstable mass fraction $f_G$ in turn would continue to drop to a level 
of a few percent.

\subsection{Intermediate Density Case}

For the intermediate-density model, we take a box size $L=5$~pc, 
a mean gas density $\rho_0=1.67\times10^{-24}$~g~cm$^{-3}$, and 
$\Gamma_{high}/\Gamma_{low}=5$.
With this setup, the mean gas density falls onto a 
stable branch of the thermal 
equilibrium curve (within the bistable pressure regime) during the high
heating state, and it stays within the unstable density range during the low
state.
As in the low-density case, we take initial conditions at $t=5$~Myr from a 
low-resolution run, which follows gas cooling in a thermally unstable regime 
from an initial temperature of $2\times10^6$~K. 
We checked a variety of periods for the time-dependent heating rate:
10~Myr (5~Myr high state and 5~Myr low state), 
2.5~Myr (0.2 + 2.3),
2~Myr (1 + 1), 
1~Myr (0.2 + 0.8),
and 1~Myr (0.5 + 0.5).
In all the cases, we find results similar to the low-density 
case in terms of kinetic energy support, with
$E_k^{max}\approx0.05E_{th}^{max}$.
The conclusions concerning Mach number, vorticity, and density variance
evolution are essentially the same as in the low-density case.

As far as the mass fractions are concerned, we find essentially more unstable
gas on the average (e.g., $f_G\approx29$\% in our experiment with a period of 
1~Myr (0.5+0.5), averaged over five periods).
The reason for this is twofold: first, the cold phase is always present and
therefore so is the unstable interphase interface 
($f_H$ saturates at $\sim55$\% in high state and at $\sim88$\% in low state);
second, the abrupt transition to the low state effectively populates 
the unstable regime at the expense of the warm stable phase, completely
consuming it until this gas reexpands to join the warm stable phase of the
low state that has a lower density than in the high state.
The bulk of this thermally unstable gas is then slowly ($\sim0.3$~Myr) 
redistributed between the cold (68\%) and warm (18\%) stable phases as the
turbulent relaxation proceeds.
In contrast to the low-density case, cores of some of the individual cold 
``clouds'' can survive several heating events without being dissolved.
In addition, new cold clouds are forming by dynamic compressions in
the disturbed warm phase ($\Delta P/P\sim10$) following transitions to the 
high state.
This mechanism, described by \citet{hennebelle.99}, is inefficient in our
low-density simulations.

\section{Discussion}
Our model is limited in many respects. 
First, it does not include effects of other driving mechanisms that can 
operate in parallel to thermal forcing.
Second, by not including magnetic fields in the simulations, our description 
of ISM dynamics is incomplete.
Third, the model does not have self-regulation, linking production of the cold 
phase to the heating rate via star formation efficiency \citep{parravano88}. 
Feedback could probably damp the oscillations occurring in response to the
time variations of the heating rate introduced ``by hand,'' but this should 
be demonstrated by an explicit calculation.
Fourth, the grain photoelectric heating rate is not simply proportional 
to gas density. 
This would slightly modify the shape of the thermal equilibrium curve as 
it shifts from a low to a high state.
Finally, the equation of state and the cooling function that we use do 
not take into 
account dynamics of ionization, recombination, formation of molecules, etc.,
which are decoupled from hydrodynamics in our treatment.
This could overestimate the minimum temperature of the cold phase and thus
underestimate its density and the rms Mach number.

Nonetheless, our main conclusion that time-dependent heating supports 
turbulence in thermally unstable multiphase ISM remains valid in 
spite of the above-mentioned model simplifications.
Thermal forcing can potentially excite turbulent oscillations on 
a wide range of length scales because of the penetrative nature of FUV 
radiation. 
The level of thermally sustained turbulence as a function of length scale at 
a given point within the Galactic disk is controlled by the local mean gas 
density and by the power spectrum of incident FUV flux time variations.
It will share the pattern of spatial inhomogeneity of FUV energy density 
within the disk.

The thermally unstable gas mass fractions in our low- and intermediate-density
models ($f_G\sim 10\%-30$\%) are in reasonable agreement with measurements
by \citet{heiles.02} as well as the fraction and morphology of the density 
distribution of the cold neutral phase (see \citet{vazquez-semadeni...02} 
for a review).
Both these results imply a substantial level of turbulence in the ISM.
We suggest time-dependent heating as one possible mechanism for driving this
turbulence.

\acknowledgments

This work was partially supported by PACI computer grant MCA98N020N
and utilized computing resources provided by the San Diego Supercomputer Center
as well as by the University of Illinois at Urbana-Champaign.

\clearpage 

\end{document}